\pdfoutput=1
\PassOptionsToPackage{unicode}{hyperref}
\PassOptionsToPackage{hyphens}{url}
\PassOptionsToPackage{dvipsnames,svgnames*,x11names*}{xcolor}
\documentclass[
  12pt,
  letterpaper,
]{article}
\usepackage{lmodern}
\usepackage{amssymb,amsmath}
\usepackage{ifxetex,ifluatex}
\ifnum 0\ifxetex 1\fi\ifluatex 1\fi=0 
  \usepackage[T1]{fontenc}
  \usepackage[utf8]{inputenc}
  \usepackage{textcomp} 
\else 
  \usepackage{unicode-math}
  \defaultfontfeatures{Scale=MatchLowercase}
  \defaultfontfeatures[\rmfamily]{Ligatures=TeX,Scale=1}
\fi
\IfFileExists{upquote.sty}{\usepackage{upquote}}{}
\IfFileExists{microtype.sty}{
  \usepackage[]{microtype}
  \UseMicrotypeSet[protrusion]{basicmath} 
}{}
\makeatletter
\@ifundefined{KOMAClassName}{
  \IfFileExists{parskip.sty}{%
    \usepackage{parskip}
  }{
    \setlength{\parindent}{0pt}
    \setlength{\parskip}{6pt plus 2pt minus 1pt}}
}{
  \KOMAoptions{parskip=half}}
\makeatother
\usepackage{xcolor}
\IfFileExists{xurl.sty}{\usepackage{xurl}}{} 
\IfFileExists{bookmark.sty}{\usepackage{bookmark}}{\usepackage{hyperref}}
\hypersetup{
  colorlinks=true,
  linkcolor=black,
  filecolor=Maroon,
  citecolor=NavyBlue,
  urlcolor=NavyBlue,
  pdfcreator={LaTeX via pandoc}}
\usepackage{longtable,booktabs}
\usepackage{etoolbox}
\makeatletter
\patchcmd\longtable{\par}{\if@noskipsec\mbox{}\fi\par}{}{}
\makeatother
\IfFileExists{footnotehyper.sty}{\usepackage{footnotehyper}}{\usepackage{footnote}}
\makesavenoteenv{longtable}
\usepackage[letterpaper,margin=1in]{geometry}

\usepackage{textcomp}
\ifPDFTeX\usepackage{mathptmx}\fi
\usepackage{iftex}
\ifPDFTeX\usepackage[T1]{fontenc}\fi
\usepackage{microtype}

\usepackage{setspace}
\usepackage{booktabs}
\usepackage{titlesec}
\usepackage{titling}
\usepackage{csquotes}
\usepackage{enumitem}

\titleformat{\section}
  {\normalfont\large\bfseries}{}{0pt}{}
\titlespacing*{\section}{0pt}{1.6ex plus 0.5ex minus 0.3ex}{0.9ex plus 0.2ex}

\titleformat{\subsection}
  {\normalfont\normalsize\bfseries}{}{0pt}{}
\titlespacing*{\subsection}{0pt}{1.2ex plus 0.4ex minus 0.2ex}{0.6ex plus 0.1ex}

\titleformat{\subsubsection}
  {\normalfont\normalsize\itshape}{}{0pt}{}

\pretitle{\begin{center}\LARGE\bfseries}
\posttitle{\par\end{center}\vspace{0.5em}}
\preauthor{\begin{center}\large}
\postauthor{\par\end{center}}
\predate{\begin{center}\normalsize\itshape}
\postdate{\par\end{center}\vspace{1.0em}}

\renewenvironment{abstract}%
  {\begin{center}\textbf{\normalsize Abstract}\end{center}%
   \begin{quote}\small\noindent\ignorespaces}%
  {\end{quote}\vspace{1em}}

\newenvironment{hangingrefs}
  {\par\begingroup
    \setlength{\parindent}{-1.5em}%
    \setlength{\leftskip}{1.5em}%
    \setlength{\parskip}{0.45em plus 0.1em minus 0.1em}%
    \noindent\ignorespaces}
  {\par\endgroup}

\author{}
\date{}

\begin{document}

\title{The Satoshi Overhang \\[0.25em] \large Why the Bear Case is Bounded}
\author{Karl T. Ulrich \\ \normalsize The Wharton School \\ \normalsize University of Pennsylvania \\ \normalsize \texttt{ulrich@wharton.upenn.edu}}
\date{Revision of June 16, 2026}
\maketitle

\begin{abstract}

Renewed attention to the identity of Bitcoin's pseudonymous creator has
revived an old worry: that the roughly 1.148 million BTC mined by
Satoshi and never moved represent a major tail risk for bitcoin. This
paper argues that the worry is overstated. The mechanical downside of
selling the position is bounded well below the feared collapse, and the
outcomes most consistent with sixteen years of observed behavior are not
bearish for bitcoin's effective supply. We analyze the position in two
ways. First, we model the case of a purely financial holder. Multiple
sale scenarios, checked against both a square-root-law estimate and the
historical record of large sales, suggest that bitcoin's current market
liquidity could absorb a patient multi-year sale with a cumulative price
impact centered around 10 to 13 percent relative to a no-sale case. The
same arithmetic also links the downside from a surprise sale to the
upside from a confirmed burn: both are bounded by the same
effective-supply adjustment, so the doom case and the burn-rally case
cannot both be large. Second, we consider the preferences implied by the
sixteen-year record. Ideological restraint, privacy, already having
enough, and preserving the myth all point toward further dormancy,
permanent loss of access, or a deliberate burn. A sale or an act of
sabotage remains possible, but the record supports it less strongly.
Under both approaches, the mechanical bear case is bounded, and the
likeliest outcomes are neutral to mildly positive for bitcoin's
effective supply. The argument does not rule out transient overshoot or
leverage-driven amplification; it bounds the durable repricing the coins
themselves can cause.



\end{abstract}

\hypertarget{introduction-and-puzzle-statement}{%
\subsection{1. Introduction and puzzle
statement}\label{introduction-and-puzzle-statement}}

Satoshi Nakamoto mined the first large share of bitcoin's money supply
during 2009 and the first half of 2010, stopped posting publicly in
December 2010, and last communicated privately in April 2011 (Nakamoto,
2008; Bradbury, 2014).\footnote{Throughout this paper, ``Satoshi'' and
  ``Nakamoto'' mean the pseudonymous creator and the associated Patoshi
  mining cluster; they do not mean the listed bitcoin-treasury company
  Nakamoto Inc.~(Nasdaq: NAKA), which shares the name but has nothing to
  do with the holder studied here.} The coins from that early mining,
about 1.148 million BTC, have never moved. Sergio Demian Lerner
identified them from the Patoshi (a blend of ``pattern'' and
``Satoshi'') pattern in the coinbase transactions of the first roughly
36,000 blocks (Lerner, 2013, 2020). Bitcoin's total mined supply is now
about 20.01 million BTC (Blockchain.com, 2026). Estimates of permanently
lost coins range from 2.78 to 3.79 million BTC in the most-cited
segmentation study (Chainalysis, reported in Roberts and Rapp, 2017) up
to substantially higher dormancy-based figures used in on-chain research
(Ambrosia, Dorrell and Stockwell, 2024). Satoshi's coins are therefore
about 5.7 percent of mined supply and roughly 7.6 percent of freely
circulating supply, defined as mined supply minus lost coins minus the
position itself (Appendix A.1).

The market has long treated this concentration as a tail risk, and the
reasoning is intuitive: a sale could overwhelm normal liquidity, and the
seller's identity could itself cause panic. The usual conclusion is that
Satoshi's coins impose a standing discount on bitcoin's value, one that
should lift only if the coins are confirmed gone for good and worsen on
any sign that they are moving. We treat this as the view to be tested,
not as a premise. Whether large holdings create lasting discounts is
still debated even for public stocks. Markets apply no agreed discount
to founder stakes of similar relative size; Tesla under Elon Musk is the
clearest current example. Large, disclosed founder sales can cause real
drawdowns, but those are usually bounded and at least partly temporary,
not catastrophic. The Satoshi position differs from a corporate
blockholding in ways that cut both ways. There is no disclosure regime,
no Rule 10b5-1 machinery, and no cash-flow floor under the price, all of
which strengthen the information channel. But there is also no company
whose control or strategy a sale would change, and sixteen years of
verifiable dormancy is a behavioral record no corporate blockholder
offers. Section 3.4 makes the word ``discount'' precise: it is the
market's residual probability of an eventual sale, multiplied by a total
spread that the absorption arithmetic fixes, and the paper's claims are
written to hold for every value of that probability.

The immediate occasion of this paper is a wave of investigative
journalism, most prominently Carreyrou (2026), naming Adam Back as the
leading candidate for Satoshi. The market's reflex is to read a more
concrete, nameable holder as more tail risk: if you can name the holder,
you can reach, coerce, tax, or persuade the holder to sell. We argue
this reflex overweights the downside, and the market's own response to
the reporting, essentially nothing on the day of publication (Section
3.2), is early evidence for that. Two fears drive the tail-risk view,
and neither holds up. The first is mechanical: that 1.148 million BTC is
simply too much for the market to absorb. Bitcoin's current liquidity
and execution capacity say otherwise; a patient multi-year sale would
cause a bounded mechanical price impact far below the collapse the
tail-risk view implies (Appendix A). The second is behavioral: that the
holder will actually run such a sale. That is hard to square with
sixteen years of revealed preference. On both counts, the plausible
outcomes for the position are neutral to mildly positive for bitcoin.
The mechanical bear case is bounded, and the likeliest outcomes are not
bearish.

This is a conceptual paper, not an empirical study. It combines stylized
market-microstructure reasoning, a scenario-based bound on mechanical
impact (cross-checked against the empirical square-root law of price
impact), event-based anchors from past large bitcoin sales,
revealed-preference inference from sixteen years of holder behavior, and
a comparative look at the cryptographic and legal tools available for
inheritance and destruction. It does not try to identify Satoshi, to
measure the market's implied probabilities, or to recommend trades. The
paper focuses on the structure of the disposition problem and the
ordering of likely outcomes.

A paper that mixes arithmetic, empirical inference, and reasoned
judgment risks letting formal language pass judgment off as fact. To
prevent that, we label the standing of each kind of claim and use the
labels consistently. \emph{Derived} claims follow logically or
arithmetically from stated assumptions; they are certain given those
assumptions and worth no more than the assumptions are. The Track 1
absorption bound (Appendix A) and the consistency-ledger identity
(Section 3.4) are derived, but they differ in an important way: the
ledger identity holds exactly, while the bound's inputs (demand
elasticity, the lost-coins adjustment, and execution friction) are
illustrative and deliberately uncalibrated, so the bound is a structured
upper bound under stated inputs, not a measurement. \emph{Inferred}
claims are read from evidence and stated with their uncertainty; the
main one here is negative, drawn from the sixteen-year record: pure
wealth-maximization is a weak prior for this holder. \emph{Judgment}
claims are the author's assessments where the evidence runs out; the
ordering of outcomes in Section 4 and their relative weight are
judgment, offered as a reasoned ranking and labeled as such, not as a
measurement or a probability. Where we call something derived, the
reader can check the derivation; where inferred, weigh the evidence;
where judgment, disagree without contradicting any finding.

The paper proceeds as follows. Section 2 explains why people came to
fear that a Satoshi sale could crash the price, and why that fear, drawn
by analogy from controlling-shareholder and artist-estate cases, is
weaker for bitcoin than it first appears. Section 3 takes the holder who
cares only about money, derives an upper bound on the mechanical bear
case, and closes with a ledger that keeps the paper's later claims about
what the market already prices mutually consistent. Section 4 takes a
holder whose profile matches the sixteen-year record and maps the range
of outcomes instead of picking one. Section 5 turns to estate planning
and argues that cryptographic inheritance tools may beat conventional
trusts on the costs this holder seems to care about. Section 6 draws out
the market implications. Section 7 places the problem in the wider class
of sales whose value depends on the act of selling. Appendix A gives the
scenarios behind Track 1; Appendix B gives the mechanics of a provable
burn.

The closest prior observation we know of is a practitioner one: Danny
Bradbury's 2014 \emph{CoinDesk} piece, ``How Dangerous is Satoshi
Nakamoto?'', in which Gavin Andresen and Sergio Lerner discussed the
possibility that Satoshi might burn the coins, with Lerner noting that
``if he did burn them, the market reaction would be terribly bullish''
(Bradbury, 2014). That remark was casual. It did not model the holder's
decision, bound the mechanical bear case, address estate planning, or
use the revealed-preference evidence. This paper takes up those tasks.

\textbf{Related literature.} Three strands of academic work bear on the
question, though none addresses it as posed here. On Track 1's
mechanical channel, the closest empirical neighbor is Ante and Fiedler
(2021), whose event study of 2,132 transfers of at least 500 BTC over
2018 to 2019 finds that the short-run price reaction to large on-chain
moves depends on transfer size and on the market's read of motive, not
on size alone. That fits the temporary-versus-permanent split used here,
but it measures reactions over hours and days to ordinary large
transfers, not the multi-year sale of a founder-scale position, and it
gives no cumulative bound. On concentration, Makarov and Schoar (2021)
use blockchain data to show that bitcoin holdings are highly
concentrated among a few large holders and intermediaries; their subject
is the active market, and they do not address what becomes of the
earliest dormant coins, so the founder-overhang question remains open in
their account. On effective supply, Ambrosia, Dorrell and Stockwell
(2024) give peer-reviewed estimates of active versus lost bitcoin that
support, rather than compete with, the lost-coins adjustment of Appendix
A.1. Alongside these sit the microstructure foundations the paper uses
directly (Kyle, 1985; Almgren and Chriss, 2001; Donier and Bonart,
2015), the blockholder and artist-estate work behind its analogies
(Barclay and Holderness, 1989; Holderness, 2003; on the blockage
discount, Center for Art Law, 2018), and the event-study methods cited
in Section 3.2 (Andersen, Bollerslev, Diebold and Vega, 2003; Bernard
and Thomas, 1989, 1990). What none of this work does, and what this
paper attempts, is to put a mechanical absorption bound, a
revealed-preference reading of sixteen years of dormancy, and a
comparative look at inheritance and destruction together into one
assessment of how the Satoshi position is likely to resolve. Section 6.1
says which parts of that assessment are new and which are synthesized.

\vspace{0.6em}

\hypertarget{the-reflexive-liquidation-frame}{%
\subsection{2. The reflexive-liquidation
frame}\label{the-reflexive-liquidation-frame}}

Standard microstructure models (Kyle, 1985; Almgren and Chriss, 2001;
Said, 2022) split the cost of selling a large position into two parts: a
temporary impact, the price concession needed to find buyers now, and a
permanent impact, the information the market reads from the order flow.
A patient seller can shrink the temporary part by spreading trades over
time and soften the permanent part by trading when information asymmetry
is low. Almgren and Chriss give a closed-form trade-off between
volatility risk and execution cost. In these models the sale value of a
large block is below its mark-to-market value, but the shortfall is
bounded and grows smoothly with size.

Two caveats. First, Almgren and Chriss is a stylized planning model: it
treats impact as fixed, parametric, and stable, whereas bitcoin's actual
impact follows the square-root law documented by Donier and Bonart
(2015) and depends on the order book and on what the market infers about
the seller's identity and intent, which is exactly the channel at issue
here. We therefore use the model only for its split of temporary and
permanent impact and its volatility-cost trade-off; none of the numbers
in Section 3 or Appendix A rely on its functional forms.

The Satoshi problem has usually been seen as going beyond this standard
picture. The extra ingredient, in the usual reading, is the collapse
that follows once the seller's identity is known. Bitcoin has no cash
flows and no intrinsic floor; its price rests on beliefs about future
scarcity, future adoption, and the intentions of its largest holder. Any
move from a known Patoshi address would be spotted within minutes and be
global news within hours, and, in the usual reading, would wipe out
whatever part of the price assumes Satoshi's coins stay dormant forever.

This is ``reflexive'' in George Soros's sense (1987, 2013): a two-way
loop in which expectations and fundamentals each move the other. Prices
are not passive readings of independent fundamentals; they help shape
the fundamentals they are supposed to measure.

Two analogies from traditional finance carry the idea to the Satoshi
case. The first is the controlling founder whose stake exceeds the
public float. The blockholder literature (Barclay and Holderness, 1989;
Holderness, 2003) finds that in such cases the information in a
founder's trading, a signaling problem in the sense of Spence (1973),
often matters more than its mechanical supply effect: the price reaction
is not just proportional to the float sold, it reflects what the market
infers about the founder's changed view of the firm. The usual fixes,
charitable transfers combined with 10b5-1 structured secondary offerings
(17 C.F.R. § 240.10b5-1), work by separating the mechanical part of a
sale from its informational part, keeping the first while defusing the
second. The Patoshi position has the same shape: mechanical absorption
is manageable under any plausible depth assumption, so the weight of the
problem falls on the information channel.

The second analogy is the estate of a dead artist holding a large
inventory whose forced, simultaneous sale would depress prices per work.
U.S. tax law recognizes this as the ``blockage discount,'' sustained at
37 percent in \emph{Estate of David Smith v. Commissioner}, 57 T.C. 650
(1972), aff'd 510 F.2d 479 (2d Cir. 1975), and at an effective 37
percent in \emph{Estate of O'Keeffe v. Commissioner}, T.C. Memo 1992-210
(Center for Art Law, 2018). This case fits the Patoshi case better than
the founder case in one respect: the originator is no longer producing,
so the market must judge scarcity without reference to ongoing output.
The standard fix, staged sale through a foundation or estate over
decades, is a direct analogue to Track 1's patient liquidation, and it
shows the problem is tractable whenever the holder, or a successor, can
spread the sale over time.

These analogies frame the concept, but they do not by themselves make
the overhang a tail risk for bitcoin. Sections 3 and 4 show that for
bitcoin in 2026 patient execution makes the mechanical absorption of
Satoshi's position a manageable constraint, and that the information
channel, though it cannot be forecast in advance, is bounded in its
mechanical part and has been weak in the nearest recorded events.

\vspace{0.6em}

\hypertarget{track-1-the-wealth-maximizers-problem}{%
\subsection{3. Track 1: The wealth-maximizer's
problem}\label{track-1-the-wealth-maximizers-problem}}

This section studies the holder of the 1.148 million BTC Patoshi
position whose only goal is personal wealth. The holder is otherwise
rational, informed, and strategic, with no ideological, legacy, or
privacy aims beyond what execution requires. Track 1 bounds the
mechanical downside for bitcoin if the real holder did happen to be a
pure wealth-maximizer.

\hypertarget{market-depth-and-absorption-arithmetic}{%
\subsubsection{3.1 Market depth and absorption
arithmetic}\label{market-depth-and-absorption-arithmetic}}

Satoshi's position is about 5.7 percent of mined supply and about 7.6
percent of freely circulating supply after the lost-coins adjustment of
Appendix A.1. At a reference price of 80,000 USD per BTC (roughly the
late-April 2026 level; Appendix A.1), the position is worth about 92
billion dollars. A patient holder running a decade-long OTC program
would sell about 115,000 BTC per year, roughly 315 BTC per day in a
market that trades around the clock. Global bitcoin spot and derivatives
volume runs in the tens of billions of dollars per day, with real
(non-wash) spot volume plausibly 10 to 20 billion. A 25-million-dollar
daily OTC flow is 0.1 to 0.25 percent of real daily volume, the kind of
background flow a mid-sized institutional desk handles routinely; even
under a steep discount to reported volumes of the kind Bitwise Asset
Management documented for the SEC in 2019, the participation rate stays
in the low single digits (Appendix A.1).

Against the heuristic elasticity range of Appendix A, roughly 0.3 (very
inelastic) to 1.5 (closer to equities), selling the full 7.6 percent of
float implies a static, partial-equilibrium cumulative price impact of
about 5 to 22 percent relative to the no-sale case, depending on
elasticity. Appendix A works through three scenarios, conservative,
base, and aggressive, each with stated assumptions about pace,
participation, elasticity, execution quality, and demand growth. The
central scenario sits near 10 percent permanent impact, 12 to 13 percent
once execution friction is added; the aggressive scenario reaches about
25 to 27 percent under low elasticity and uneven execution. An
independent check using the empirical square-root law of price impact,
which shares no parameters with the elasticity arithmetic, lands in the
same range (Appendix A.8).

Empirical episodes bound the \emph{temporary} part of impact under
different execution styles, while the elasticity arithmetic addresses
the \emph{permanent} part. Between June 19 and July 12, 2024,
authorities in the German state of Saxony sold 49,858 BTC seized in a
criminal case, in publicly tracked tranches over twenty-three days, and
bitcoin fell roughly 13 to 15 percent over the episode, a window that
also held the announced Mt. Gox distribution of about 140,000 BTC and a
forced unwind of leveraged longs (CoinDesk, 2024). The price began
recovering within days of the final tranche and was back to its
pre-episode level within weeks. A sale of about 0.3 percent of float
whose price effect mostly reverses has a large temporary part and a
permanent part near zero; read as permanent, it would imply a demand
elasticity an order of magnitude below any proposed value, and the
recovery is the evidence against that reading. By contrast, the U.S.
Marshals auctions of Silk Road coins, less publicized and sold straight
to institutional buyers, drew muted or even positive responses; the June
2014 auction of about 29,656 BTC, won in full by a single bidder, was
followed by a price rise. This evidence suggests that the price effect
of a large on-chain transfer depends on the motive the market infers
behind it, not on the amount moved alone (Ante and Fiedler, 2021).

The most direct empirical anchor is the most recent. In July 2025 a
holder dormant since April 2011, the month of Satoshi's last known
communication (the private emails to Hearn and Andresen; the last public
forum post was December 2010), moved about 80,000 BTC and sold the whole
position, roughly nine billion dollars, through a single institutional
OTC desk in a disclosed estate-planning transaction. The price fell a
few percent and recovered within days (CoinDesk, 2025). This is the
closest thing on record to a Patoshi sale: same dormancy vintage,
one-fourteenth the size, run through exactly the channel Track 1
describes. As an observed event, it is a firmer anchor than the
unaudited practitioner estimates of OTC impact that such arguments
usually rest on, and it fits the peer-reviewed finding that impact from
flow the market reads as uninformed decays almost completely over time
(Donier and Bonart, 2015).

A reasonable point estimate for the permanent cumulative impact of a
patient ten-year Satoshi sale, under disciplined execution and continued
demand growth, is therefore in the mid-single to low-double digits, with
the aggressive assumptions reaching the mid-twenties. Even that upper
figure is far below the share-of-current-price loss the existential-tail
reading implies.

\hypertarget{the-information-channel}{%
\subsubsection{3.2 The information
channel}\label{the-information-channel}}

Section 2's reflexivity argument holds that identity revelation, not
mechanical volume, would drive a catastrophic price drop. Judging it
means separating what can be bounded from what can only be guessed. The
event-study literature shows that only the unexpected part of news moves
prices (Andersen, Bollerslev, Diebold and Vega, 2003), and that
reactions to fundamental news run the gamut, from delayed under-reaction
and drift (Bernard and Thomas, 1989, 1990) to overreaction and reversal.
For a one-of-a-kind event with no precedent, an authenticated Satoshi
action, the market's prior is not observable, so the surprise cannot be
known in advance. We therefore make no point forecast of the size, sign,
or persistence of the reaction. The limit cuts both ways: it equally
forbids the confident existential forecast the tail-risk view depends
on. What it leaves us is the bounded mechanical part (Section 3.4) and
the record of nearby events.

Two natural experiments are on record. On April 8, 2026, the most
prominent identity reporting in bitcoin's history named a specific,
living, reachable person as the leading Satoshi candidate (Carreyrou,
2026; CNBC, 2026). Bitcoin was essentially flat that day, with steady
derivatives open interest and no liquidation cascade. Either the market
discounted the identification, which the named person denied and whose
stylometric support the investigation's own consultant called
inconclusive, or marginal flows simply do not respond to founder news;
both readings contradict the claim that attention to identity is itself
existential. The second experiment is the July 2025 sale described in
Section 3.1, which paired a reveal (a fourteen-year-dormant wallet of
Satoshi's own vintage moving) with an actual sale and produced a
low-single-digit, fully reverting response. Neither event measures the
response to an authenticated Satoshi action: the first was an
unconfirmed identification, the second was one-fourteenth the size, and
two points are not a distribution. But they anchor the market's observed
sensitivity in the two nearest available cases, and in both the
sensitivity was small and short-lived. A catastrophic interpretation
therefore requires a sharp discontinuity between these analogues and an
authenticated Patoshi event.

One conditional channel deserves explicit mention: institutional
compliance. ETF sponsors, regulated custodians, and corporate treasuries
work under compliance rules that ignore founder news for ordinary
allocation decisions but do track regulatory exposure. If a named
Satoshi is a national or resident of a country under international
sanctions, has an unresolved tax liability, or faces active litigation,
compliance teams may pause new allocations pending legal review. The
effect is temporary and symmetric: allocations resume once review ends.
Its size and length depend on the specific identity and exposure, not on
the bare fact of a reveal.

A second channel runs the other way. For an asset with no cash-flow
floor, both demand elasticity and the float depend on the narrative: a
Patoshi sale that dented the upside story would pull demand in and push
more supply onto the market at the same time, so effective elasticity is
lowest exactly when a sale is most likely. The mechanical bound of
Section 3.1 is therefore a bound on the supply arithmetic, not on total
repricing in a bad narrative; Appendix A.2 shows the amplifying and
dampening adjustments side by side.

The useful distinction the bound supports is between an existential and
a material drawdown. Existential, the implicit claim of the tail
framing, means permanent loss of a large share of value. Material means
a path on the order of the mechanical bound plus a transient overshoot
of unknown size, within bitcoin's ordinary yearly volatility and
historically followed by recovery. For unleveraged holders the
distinction is decisive. For leveraged ones it is not: funds and
digital-asset-treasury vehicles whose financing is marked to net asset
value can be forced to sell into a material drawdown, and the 2025-26
stress among treasury vehicles trading below the value of their
holdings, with covenant-driven sales, shows the propagation machinery is
real. Section 7 returns to it. What the record does not support is the
stronger claim that founder news on its own, without these leverage
channels, commands a large lasting repricing.

In sum, the lasting repricing from any Patoshi event is a bounded total
spread times the market's revision in the probability of an eventual
sale (Section 3.4). The transient part cannot be forecast, and we do not
forecast it; and the two nearest events on record produced reactions
that were small, short-lived, or both. The paper bounds the durable
supply-driven component of the reaction rather than forecasting the
total price path: whatever the probability revision turns out to be, the
durable repricing is capped by the absorption arithmetic, and the
direction of the response to a confirmed burn or non-recovery follows
from the supply identity (Section 6.2), not from a prediction of its
size or path. Resting on that arithmetic and on the revealed-preference
evidence, the argument needs no forecast of how the market would react
to news.

\hypertarget{optimal-disposition-under-pure-wealth-maximization}{%
\subsubsection{3.3 Optimal disposition under pure wealth
maximization}\label{optimal-disposition-under-pure-wealth-maximization}}

Combining the absorption arithmetic of 3.1 with the information analysis
of 3.2, a wealth-maximizer's problem reduces to choosing among versions
of patient sale. The holder could announce the program in advance,
keeping pricing credible at the cost of the initial information shock.
The holder could sell quietly through many OTC desks, accepting the
operational cost and the risk of eventual identification. Or the holder
could cap the pace of an announced program in code, using CLTV timelocks
(BIP 65, activated December 2015; Todd, 2014) or CSV timelocks (BIP 112,
activated May 2016; BtcDrak, Friedenbach, and Lombrozo, 2015), turning a
signed promise from cheap talk into a protocol-enforced limit. This
announced-and-capped version is like a Rule 10b5-1 plan, but stronger,
because the signature from the Patoshi keys cannot be forged and the
timelock is enforced by the protocol rather than by a rule.

A further version reverses the order: the holder first builds a short
position in bitcoin derivatives sized against the coins, then moves the
coins and closes both legs together with a time-weighted algorithm. The
appeal is that the repricing triggered by detection lands on the
derivative counterparties rather than on unhedged coins, locking in
pre-reveal prices; for the holder, the information shock becomes roughly
a wash. Two facts bound the strategy. First, large short positions are
visible: in October 2025 a single actor's leveraged bitcoin and ether
perpetual shorts of more than one billion dollars on a public on-chain
venue were spotted by surveillance within hours of being opened, so the
signature of such a program would not stay hidden. Second, scale binds.
Total bitcoin futures open interest across venues ran about 30 to 45
billion dollars in the first half of 2026, with CME about 10 billion of
that, so fully hedging the roughly 92-billion-dollar position would mean
becoming a short side two to three times the size of the entire futures
market, backed by tens of billions in outside collateral the Track 1
holder does not have, all while funding rates and open interest
broadcast the program. The realistic version is a partial hedge that
trims the temporary and overshoot costs the seller bears. The permanent
part is unchanged: counterparties reprice it on detection, ahead of the
flow. This version actually strengthens the section's conclusion, since
it pushes the wealth-maximizer's outcome closer to the mechanical bound.
We take no view on where such a program would stand under derivatives
anti-manipulation law.

In every version, the wealth-maximizer nets on the order of 50 to 100
billion dollars over the sale period, against roughly zero from an
outright burn. For a holder who cares only about wealth, burning is
strictly worse than patient sale on every horizon. So the paper has
nothing to tell such a holder except that the best strategy is patient
sale, and the market should not price that outcome as an existential
tail risk. How strongly the claims of 3.1 and 3.2, and the market
implications of Section 6, can be stated all turn on one unobserved
quantity, the share of the Patoshi position that the current price
already treats as permanently out of float.

\hypertarget{a-consistency-ledger-for-the-dormancy-premium}{%
\subsubsection{3.4 A consistency ledger for the dormancy
premium}\label{a-consistency-ledger-for-the-dormancy-premium}}

This section develops a ledger for that unobserved quantity. Writing the
bookkeeping out lets the paper's separate claims be checked against each
other, and shows they hold together for every admissible value of that
share. The formal version, with its constant-elasticity demand schedule,
is in Appendix A.7. In words, the argument runs as follows.

Confirmation that the coins are gone for good, and confirmation that
they have all been sold, mark the two ends of a price interval. Ignoring
any unresolved-state discount, the current price sits inside this
interval, at a point set by the share the market already prices as
eventually returning to float. With such a discount, the current price
sits below that interior point, and in principle can fall below the
sale-confirmed endpoint. The width of that interval is fixed by the
absorption arithmetic of Appendix A, not by the unknown share: about 11
percent at the central elasticity, narrowing to about 5 percent in the
elastic case and widening to about 27 percent in the inelastic case. We
do not try to estimate the share; the point of the ledger is that what
follows holds whatever it is.

First, the gain from a confirmed burn and the loss from a confirmed but
unexpected sale are complementary: each is a slice of that one fixed
interval, so they cannot both be large. The larger the share the market
has already priced as permanently gone, the larger the sale downside and
the smaller the burn upside; the two always sum to the same fixed
interval. Second, the Appendix A figures are the full-surprise case, in
which nothing was priced in, and so are upper bounds on the unexpected
part of a sale. Third, a sign-of-life reveal with no coin movement
reprices only the revision in that share which the reveal itself causes,
and the revealed-preference evidence of Section 4 is the reason to
expect that revision to be small.

The Section 6 claim about the response to a confirmed burn has to be
read as proportional to the residual sale probability still embedded in
the price: if on-chain practice has already absorbed most of the
dormancy discount, the supply part of the burn response is positive but
moderate, with any additional effect coming from retiring the
uncertainty premium. And the existential-tail view now needs either a
demand elasticity far below any proposed value or a reflexive
demand-side amplification of the kind discussed in 3.2 and Section 7.
The doom case and the burn-rally case are complementary shares of the
same bounded total, and thus both cannot comprise a majority.

\vspace{0.6em}

\hypertarget{track-2-the-observable-satoshi-profile}{%
\subsection{4. Track 2: The observable Satoshi
profile}\label{track-2-the-observable-satoshi-profile}}

This section studies a holder whose identity is unknown but whose
profile is partly readable from sixteen years of behavior. We stay
agnostic about whether the holder is Adam Back, another early cypherpunk
who has not drawn media attention, or someone no outsider has
identified. Unlike Track 1, which gave a number, Track 2 maps the range
of possibilities: the motivations consistent with the record, and the
outcome each would justify. We do not pick a single most-likely outcome.

\hypertarget{revealed-preference-what-sixteen-years-rule-in-and-rule-out}{%
\subsubsection{4.1 Revealed preference: what sixteen years rule in and
rule
out}\label{revealed-preference-what-sixteen-years-rule-in-and-rule-out}}

Sixteen years of Patoshi dormancy is a behavioral dataset. A holder who
valued more money above all else would not likely have stayed silent
through four bull markets (2013, 2017, 2021, 2024), each a good chance
to exit. Even a clumsy partial sale in any of them would have raised
sums large enough to swamp almost any reasonable spending or bequest
plan. The holder did none of this. The prior on pure
wealth-maximization, the premise of Track 1, is therefore empirically
weak.

The record also narrows the holder's capabilities. Satoshi's writing in
the whitepaper, the early BitcoinTalk posts, and the reference client's
source code shows command of applied cryptography, distributed systems,
and economic design that points to a small subset of the cypherpunk
community active from 2007 to 2010. The operational security held across
two years of pseudonymous communication, the clean exit in April 2011,
and the sixteen years of silence since show discipline that narrows the
profile further. This matters later: a holder of this competence can
execute any disposition, including the cryptographic inheritance and
destruction arrangements of Section 5, cleanly and without help.

So the record establishes the holder's capability, the cryptographic
skill and operational discipline to carry out any disposition, and it
makes pure wealth-maximization a weak explanation for the silence,
though dormancy by itself cannot separate deliberate restraint from lost
keys, incapacity, death, or group deadlock. What it does not establish
is the motive behind the dormancy. The whitepaper's ideological cast,
hard money, supply discipline, and disintermediation from central
institutions, hints at an orientation often associated with what is now
called the bitcoin maximalist tradition. Whether that orientation, or
some other preference, actually governs the holder's behavior is the
question the next section takes up.

\hypertarget{alternative-preference-sets-consistent-with-dormancy}{%
\subsubsection{4.2 Alternative preference sets consistent with
dormancy}\label{alternative-preference-sets-consistent-with-dormancy}}

Dormancy fits many sets of preferences, not one. Before naming likely
outcomes, we lay out the preferences that would explain the behavior we
see.

\emph{Ideological restraint.} Taking up the orientation noted in §4.1,
the holder treats bitcoin's supply discipline and the symbolic weight of
the Patoshi position as matters of principle and leaves the position
untouched as a commitment. This is the maximalist case: the view that
bitcoin alone, with its fixed supply, is the legitimate digital money,
and that its supply discipline must not be diluted by rival protocols,
contentious forks, or accommodation of centralizing intermediaries. It
is the case most of this paper focuses on.

\emph{Privacy above all.} The holder's main aim is to stay pseudonymous.
Any disposition, even a careful sale, creates detection risk that
outweighs the money. Dormancy is the lowest-risk choice.

\emph{Having enough.} The holder reached a point of having enough long
ago, through other wealth, a modest life, or both, and gains little from
more. Dormancy is the default when there is no positive reason to act.

\emph{Lost keys or incapacity.} The holder has lost the Patoshi keys, is
incapacitated, or has died with no successor plan. Dormancy is
mechanical, not chosen. A variant: Satoshi was more than one person, and
disagreement has blocked any coordinated action.

\emph{Myth preservation.} The holder judges the position's value as a
permanent, untouched reserve to be worth more than anything spending it
could buy, and keeps the myth intact.

\emph{Legal caution.} The holder sees realizing the wealth, or revealing
an identity, as inviting tax, regulatory, or criminal exposure
(money-laundering characterization, sanctions, securities treatment)
whose expected cost beats the gain.

These motivations are not mutually exclusive. Ideological restraint and
privacy likely travel together in any profile that fits the technical
record. Lost keys and incapacity look the same as the others without
positive evidence. The group-stalemate variant under ``lost keys or
incapacity'' is worth a note: a Satoshi of several people whose
preferences now diverge, and who therefore cannot agree to act, produces
the same observable pattern as a single holder's ideological restraint,
since inaction is the default either way. The analysis below leans on
the outcomes consistent with ideological restraint and privacy, but
notes that having enough, lost keys, myth preservation, legal caution,
and group stalemate are all consistent with the record and would look
much the same in the near term.

\hypertarget{the-adversarial-dead-mans-switch}{%
\subsubsection{4.3 The adversarial dead-man's
switch}\label{the-adversarial-dead-mans-switch}}

One motivation we have not yet covered is adversarial. A maximalist
holder might read later developments, the centralization of mining and
custody, regulatory capture through KYC rules, or dilution of supply
discipline through forks, as betrayals of the project, and arm a
dead-man's switch that sells or dumps the position if specified events
occur. This is cryptographically feasible and has been discussed
informally among cypherpunks.

Three things weigh against it as the dominant motive. The holder has had
cheaper, more surgical ways to signal hostility over sixteen years,
public denunciation, funding a rival protocol, a paper on protocol
correctness, and has used none. A holder who could build a hostile
switch could as easily build a constructive one, so the choice of a
hostile design would itself reveal a preference the record does not
show. And a dormant hostile switch is fragile: it must track protocol
developments for decades, resist manipulation, and never fire by
accident, whereas destroying the keys at death removes those failure
modes at the cost of the hostile capability. We do not rule the switch
out; we rank it below the non-hostile outcomes on the strength of the
record.

\hypertarget{terminal-dispositions-consistent-with-the-record}{%
\subsubsection{4.4 Terminal dispositions consistent with the
record}\label{terminal-dispositions-consistent-with-the-record}}

Three outcomes fit the motivations of 4.2 and 4.3 best. We list them in
rough order of fit with the record. The ordering is ordinal, not a
probability assignment.

\emph{Continued dormancy ending in permanent, cryptographically enforced
loss of access.} Over a remaining life of some decades, the holder stays
silent and arranges for the Patoshi keys to become permanently
unrecoverable at or near death, using one of the cryptographic tools in
Section 5. This needs no announcement, no reveal, no pre-commitment
apparatus, no legal machinery, and no change from the sixteen-year
status quo. It fits the ideological, privacy, and myth-preservation
motivations best.

\emph{A silent, unattributed burn, in whole or in part.} At a time of
the holder's choosing late in life, a transaction from the Patoshi
addresses to an OP\_RETURN output takes the position, or most of it, out
of float; Appendix B covers the mechanics and how anyone can verify
them. A burn is an action rather than an absence, so it is a little more
exposed operationally than key destruction, but pseudonymity can hold
because reading the event does not require knowing who did it. A
practically useful variant keeps a small slice for option value: sending
about 99 percent of the cluster (around 1.136 million BTC) to an
OP\_RETURN output and about 1 percent (around 11,500 BTC, roughly one
billion dollars at current prices) to a fresh non-Patoshi address looks,
at the market's resolution, the same as a full burn while keeping some
optionality. The bullish signal from a 99 percent burn is effectively
the same as from a 100 percent burn. This outcome fits ideological
restraint and myth preservation, and, through the retained slice, a
residual of having-enough or legal-caution motivation.

\emph{A hostile switch.} Programmed sale or dumping triggered by
protocol developments, as in 4.3. Ranked below the two above on the
record, but not ruled out.

The ordering is ordinal and rests on the record. It would shift if a
motivation not listed in 4.2 came to light. It is reinforced, not upset,
by evidence that the holder is incapacitated, that the keys are lost, or
that Satoshi was a group whose members now disagree: each of these makes
continued dormancy more likely as the realized outcome. We flag these
sensitivities rather than try to price them.

For readers who want the ordering carried toward the pricing
recommendation of Section 6, the bearish branch can be explored through
reader-side sensitivity cases rather than assigned a probability. Let q
be the probability that the outcome is bearish, a wealth-maximizing sale
or a hostile dump rather than dormancy ending in lost access or a burn.
For illustration, one can test three reader-chosen cases rather than
assign a probability: q = 5 to 15 percent, q = 25 to 40 percent, and q =
50 percent, corresponding to a reader who accepts the
revealed-preference argument, one who is agnostic, and one who discards
the inference from dormancy. The point of showing all three is that the
Section 6.2 recommendation does not depend on the choice. Because even
the bearish branch is bounded by the absorption arithmetic of Section
3.4, the probability-weighted downside is a fraction of an already
bounded total at every value, and the qualitative conclusion, that the
position should be priced as effectively removed supply carrying a
small, bounded residual sale risk, survives even the skeptical case.
These are stress cases for the reader, not estimates; in keeping with
Section 3.2, the paper assigns no probabilities of its own.

\hypertarget{why-the-likely-terminal-states-are-non-bearish}{%
\subsubsection{4.5 Why the likely terminal states are
non-bearish}\label{why-the-likely-terminal-states-are-non-bearish}}

The first two outcomes reach the same physical supply result: about
1.148 million BTC, or very nearly so in the retained-slice variant, is
permanently out of float. They differ in verifiability, which matters
for pricing (Section 3.4). A provable burn settles the market's residual
sale probability in one block; it is the only outcome that delivers a
discrete, fully credible confirmation. A key-destruction arrangement,
however sincere, can never be verified from outside, so confirmation
arrives only gradually, as the market updates on continued dormancy, and
part of the upside is delivered slowly and part, in principle, never.
The two are physically the same for supply and different for price.
Either way, sooner or later, the market gets a permanent subtraction of
about 5.7 percent of mined supply.

The third outcome, the hostile switch, would be bearish but fits the
record least, for the reasons in 4.3. On the record-based ordering, the
likeliest outcomes are neutral to mildly positive for bitcoin's
effective supply, not the loss the tail-risk view implies. This is not a
strong bullish base case. It is the claim that the motivations most
consistent with the record lead to outcomes that are not bearish, and
that the bearish outcome is the one requiring the largest departure from
the observed profile. The conclusion is admittedly over-determined:
dormancy, having enough, lost keys, death without a successor, group
stalemate, and a burn all map to the same non-bearish supply result, so
``not bearish'' is robust precisely because so many different
motivations converge on it. That convergence is substantive, not true by
construction, because the hostile switch of 4.3 is a genuine bearish
outcome the framework allows; what holds it down is the sixteen-year
record, not the accounting.

\vspace{0.6em}

\hypertarget{estate-planning-cryptographic-primitives-versus-conventional-trust-machinery}{%
\subsection{5. Estate planning: cryptographic primitives versus
conventional trust
machinery}\label{estate-planning-cryptographic-primitives-versus-conventional-trust-machinery}}

A standard treatment of a position this large would turn here to trusts,
tax planning, and transfer mechanisms. The problem is not hypothetical:
the July 2025 sale of more than 80,000 BTC dormant since April 2011 was
disclosed by the executing desk as part of the holder's estate planning,
a real case of a Satoshi-era holder facing exactly this question.

A conventional trust imposes three costs on this kind of holder. First,
disclosure: the settlor's identity goes to the trustee, into the trust
documents, and, depending on jurisdiction and entity structure, into
beneficial-ownership registries and other regulatory filings. Second,
counterparty exposure: the trust carries lifetime exposure to trustee
discretion, to beneficiary litigation, and to administrative processes
that can be subpoenaed or surfaced in estate proceedings. Third, process
cost: trust administration is slow and expensive next to the
cryptographic tools the holder has used for sixteen years. Most large
holders can tolerate these costs, because their identity is already
public and there is no good alternative. For a holder who has kept full
cryptographic self-custody throughout, the costs are a break from
demonstrated preferences, and cryptographic substitutes preserve the
pattern.

The obvious alternative, simply handing private keys to adult heirs, is
also a break from preference. Heirs usually lack the operational
security the holder has kept for sixteen years, and once they hold the
keys, nothing binds them to the holder's wishes about dormancy or
disclosure. A holder who has protected this position by trusting no one
is unlikely to end the program by trusting his children with bare keys.

The natural substitutes encode the holder's preferences into the
mechanism rather than handing them to a person, and several are
production-grade today. Shamir Secret Sharing (Shamir, 1979) splits a
key into m-of-n shards so that reconstruction needs a quorum; spreading
shards across heirs, trustees, and separated custodians builds in delay,
consensus, or conditional reconstruction. Multisig wallets with CLTV
(BIP 65) or CSV (BIP 112) timelocks block any unilateral move until
after events the holder specifies, a fixed date, an elapsed time, or a
protocol-observable trigger. Modern threshold-signature schemes such as
MuSig2 (Nick, Ruffing, and Seurin, 2021) and FROST (Komlo and Goldberg,
2021) produce on-chain transactions that look like ordinary
single-signature ones and allow m-of-n reconstruction without revealing
the threshold. A dead-man's switch that releases shards or sends a
predetermined transaction after a long silence is straightforward to
build on these tools. And deliberately destroying the shards at death
makes the coins unrecoverable by anyone, heirs included, turning the
bequest into a contribution to bitcoin's supply discipline.

These tools are ones the holder understands natively and trusts by
construction, and they keep pseudonymity across generations in a way
conventional trusts are less well suited to. The claim is not that
trusts are worse in general, but that for a holder with the preferences
of Section 4.1, cryptographic substitutes may win on the dimensions that
matter most here: disclosure, counterparty exposure, and process cost.

A structural point connects this to the economics of trust law. Modern
trust and estate practice exists to handle the principal-agent problem
that arises when a settlor cannot personally carry out a multi-decade,
preference-consistent plan after death. The trustee is the settlor's
agent, and the apparatus of trust law, fiduciary duty, accounting,
beneficiary standing, aligns the agent with the settlor's prior wishes.
The cryptographic tools change the problem: a timelock, a quorum of
shardholders bound only to a mechanical reconstruction rule, or a
programmed destruction instruction involves no agent to incentivize or
monitor. The settlor's preferences are written straight into the
mechanism that carries them out. The agency problem is not merely
cheaper to manage; if the mechanism is well designed, the discretionary
part is sharply reduced, though shardholders, heirs, and custodians can
remain quasi-agents at the edges. That is a stronger reason to
substitute than ordinary cost-minimization alone suggests.

One tax point is worth noting. Under U.S. Internal Revenue Code Section
1014, property held at death gets a stepped-up basis equal to its fair
market value at death (26 U.S.C. § 1014; IRS Publication 551). But
claiming the step-up means reporting the coins as part of the estate,
which breaks the pseudonymity the profile has protected for sixteen
years. A holder who places pseudonymity and supply discipline above tax
savings would forgo the step-up, and the bequest to heirs becomes the
holder's continued participation in bitcoin rather than the coins as a
spendable asset.

\vspace{0.6em}

\hypertarget{market-implications}{%
\subsection{6. Market implications}\label{market-implications}}

The two-track analysis yields two main findings. The mechanical downside
is bounded even under Track 1's purely financial assumption that the
holder cares only about money. And the outcomes most consistent with the
record are neutral to mildly positive for bitcoin's effective supply.

\hypertarget{what-the-market-may-or-may-not-already-price}{%
\subsubsection{6.1 What the market may or may not already
price}\label{what-the-market-may-or-may-not-already-price}}

Crypto-native participants are sophisticated, and none of this paper's
pieces is wholly new. Lerner's identification of the Patoshi cluster has
been public since 2013, OP\_RETURN has been standard since 2014,
Bradbury's 2014 piece raised burning informally, and the lost-coins
adjustment is routine in on-chain research. It is plausible that much of
the probability weight on permanent dormancy is already in the price
through some form of that adjustment. The April 2026 episode fits this
reading. The point is testable: the size of the dormancy discount, and
changes in it, can be estimated with event studies around large
dormant-wallet movements and credible Satoshi rumors, of which the July
2025 sale and the April 2026 reporting are the two most recent and
cleanest cases, and the short-horizon design of Ante and Fiedler (2021)
is a natural template. We do not run that test here, but the design
exists and these two episodes are its first data points. We make no
attempt to measure how much of the position the market already prices as
permanently gone, and, as in Section 3.4, the paper's claims are built
to hold for every value of it.

Three parts of the argument are less present in current practitioner
writing. The quantitative upper bound on the mechanical bear case
derived in Section 3.1 does not, as far as we can tell, appear in
published or practitioner sources. The cost case against conventional
trusts for this profile has not been made in the estate-planning or
asset-management literature. And the explicit map of Track 2's outcomes,
with its ordering and its list of alternative motivations, is, to our
knowledge, new. Whether any of this is in current prices is a separate
empirical question we do not settle.

\hypertarget{implications}{%
\subsubsection{6.2 Implications}\label{implications}}

The paper's claims are structural, not claims that the market is
mispriced. The bear-case bound of Section 3 rests on bitcoin's current
liquidity and execution capacity, and the outcome map of Section 4 rests
on the sixteen-year record. Both are observable independent of price,
and the implications below follow from that structural account rather
than from any assumption about what the market currently prices (see
also Section 6.1).

First, any verifiable event matching one of Track 2's first two
outcomes, in particular a confirmed death of a profile-matching
candidate followed by evidence of non-recovery, or a burn of the
position, would likely read as a positive signal of a confirmed supply
subtraction. The size of the mechanical response is set by the ledger of
Section 3.4: proportional to the residual sale probability still in the
price, plus the retirement of the uncertainty premium, and bounded by
the same arithmetic that bounds the mechanical bear case. The supply
component carries an unambiguous positive sign under every calibration;
the total market reaction may bring in other channels (identity, legal,
tax, or macro context), whose sign and size we do not predict. The two
events also differ in kind: a burn delivers the repricing in one block,
while non-recovery is confirmed only gradually (Section 4.5), so the
burn is the only outcome that produces a discrete event at all.

Second, practitioners do not, on the whole, hedge ``Satoshi risk'' as a
named state; they trade bitcoin volatility, to which a Patoshi event is
one contributor among macro shocks, regulatory moves, exchange failures,
and leverage unwinds. The analysis here says that contribution should be
sized as a material-drawdown scenario, on the order of the mechanical
bound plus a transient overshoot of unknown size, not as a separate
existential state worth deep out-of-the-money protection at any price.
For unleveraged holders, the material-versus-existential distinction is
decisive. For leveraged holders and treasury vehicles financed against
net asset value, a material drawdown is itself serious, and the
propagation machinery shown in the 2025-26 treasury-vehicle stress is
real; Section 7 returns to it.

Third, pricing models that treat the Patoshi coins as circulating supply
discounted by a probability of sale should be replaced by models that
treat them as effectively removed supply carrying a small residual
probability of wealth-maximizing sale, itself bounded by the absorption
arithmetic of Section 3.

\vspace{0.6em}

\hypertarget{the-broader-class-of-reflexive-liquidation-problems}{%
\subsection{7. The broader class of reflexive liquidation
problems}\label{the-broader-class-of-reflexive-liquidation-problems}}

The Satoshi case is one instance of a broad class in financial
economics: positions whose sale value is set partly or wholly by the act
of selling. The class is not new, but its members have usually been
studied separately rather than together. A preliminary taxonomy follows.

Controlling founders of listed companies. When a founder holds a stake
well above the public float, a sale both adds supply and removes the
signal of founder conviction that partly holds the price up. Barclay and
Holderness (1989) and the later blockholder literature document the
price effects. The usual fixes, charitable transfers plus structured
secondary offerings, are partial analogues to Track 1's patient sale.

Estates of dead artists. The blockage discount in U.S. tax law,
sustained at 37 percent in \emph{Estate of David Smith} (1972) and at an
effective 37 percent in \emph{Estate of O'Keeffe} (1992), recognizes
that selling a large inventory of one artist's work at once depresses
prices per work. The usual fix is staged sale through a foundation or
estate, in effect a decades-long schedule.

Insider stakes in private companies. With no public market, an early
holder's attempt to sell through secondaries can signal to new investors
that the company is overvalued, depressing both the stake's price and
the valuation of later rounds.

Leveraged treasury vehicles holding the asset itself. A 2025-26 case
closes the loop: listed digital-asset-treasury companies hold bitcoin
against narrative-sensitive capital structures, and when their market
value falls below the value of their holdings, covenant and financing
pressure can force staggered sales into weakness. A Patoshi move is one
of several possible triggers of such a cascade, and the amplification it
could cause may exceed the direct Patoshi flow bounded in Section 3.
This does not change the mechanical bound, which is about the holder's
own flow; it identifies the main path by which a material drawdown could
spread, and it belongs to this class in its own right.

The Satoshi case shares the class's structure but differs in three ways
that matter in 2026. First, bitcoin's market, unlike a single stock or a
private company's secondary market, has grown liquid enough to absorb
the full position through patient sale, which bounds Track 1's
mechanical bear case in a way the classic blockholder case is not.
Second, the holder's identity is unknown, so the information asymmetry
is not about a known person's changed circumstances but about the prior
question of whether the holder is alive, rational, and paying attention,
a different channel from anything in the blockholder literature. Third,
bitcoin offers a uniquely credible destruction option: a transaction to
an OP\_RETURN output is a voluntary, verifiable, irreversible supply
subtraction (Appendix B), and no other asset in this class gives its
holders that option.

That destruction option has a theoretical consequence. For an asset
whose value rests partly on expected supply limits, a large holder's
credible option to destroy is itself a positive contribution to the
price, whether or not it is ever used. A marginal buyer can reasonably
assign some probability to the destruction event and price expected
supply accordingly. The option's credibility rests on the verifiability
detailed in Appendix B: it is the provable burn, not key destruction,
that makes the option contract-grade. The asymmetry favors the asset:
the destruction option adds supply discipline in expectation, the sale
option adds supply in expectation, and the sale carries low probability
for the holder who actually exists under the Section 4 profile.

The broader implication is that the study of illiquid and reflexive
positions should account for voluntary-destruction options where they
exist, and treat their absence in traditional assets as a real
constraint on those holders. Bitcoin is unusual in offering one, and
that option, available to the Satoshi holder and to no comparable holder
elsewhere, is what most distinguishes this overhang from the tail risk
it is taken to be.

\vspace{0.6em}

\hypertarget{acknowledgements}{%
\subsection{Acknowledgements}\label{acknowledgements}}

I benefited from conversations with Shan Wang about the logic in the
paper. I thank three anonymous reviewers and the editor at \emph{Ledger}
for comments that materially improved the paper, in particular the
consistency ledger of Section 3.4, the derivatives-first execution
variant of Section 3.3, and the treatment of provable destruction in
Appendix B. As a matter of disclosure, my first real education about
Bitcoin was in 2017 from a friend who had a single machine in his garage
mining coin. I bought 1 BTC at that time just for fun, and I've not sold
it. That's my only financial interest in Bitcoin.

\vspace{0.6em}

\hypertarget{references}{%
\subsection{References}\label{references}}

\begin{hangingrefs}

Almgren, R., and N. Chriss (2001). ``Optimal Execution of Portfolio
Transactions.'' \emph{Journal of Risk}, 3(2): 5-39. DOI
10.21314/JOR.2001.041.

Ambrosia, M., J. Dorrell, and T. Stockwell (2024). ``Is active bitcoin
supply decreasing? An empirical analysis.'' \emph{Journal of Economics
and Finance}, 48(4): 1166-1186. DOI 10.1007/s12197-024-09691-w.

Andersen, T. G., T. Bollerslev, F. X. Diebold, and C. Vega (2003).
``Micro Effects of Macro Announcements: Real-Time Price Discovery in
Foreign Exchange.'' \emph{American Economic Review}, 93(1): 38-62. DOI
10.1257/000282803321455151.

Ante, L., and I. Fiedler (2021). ``Market reaction to large transfers on
the Bitcoin blockchain. Do size and motive matter?'' \emph{Finance
Research Letters}, 39, 101619. DOI 10.1016/j.frl.2020.101619.

Barclay, M. J., and C. G. Holderness (1989). ``Private benefits from
control of public corporations.'' \emph{Journal of Financial Economics},
25(2): 371-395. DOI 10.1016/0304-405X(89)90088-3.

Bernard, V. L., and J. K. Thomas (1989). ``Post-Earnings-Announcement
Drift: Delayed Price Response or Risk Premium?'' \emph{Journal of
Accounting Research}, 27 (Supplement): 1-36. DOI 10.2307/2491062.

Bernard, V. L., and J. K. Thomas (1990). ``Evidence that stock prices do
not fully reflect the implications of current earnings for future
earnings.'' \emph{Journal of Accounting and Economics}, 13(4): 305-340.
DOI 10.1016/0165-4101(90)90008-R.

Bitcoin Core (2025). ``Bitcoin Core 30.0 Release Notes.''
bitcoincore.org. October 2025.

Bitcoin Wiki (2024). ``OP\_RETURN.'' en.bitcoin.it/wiki/OP\_RETURN.
Accessed April 2026.

Bitwise Asset Management (2019). ``Economic and Non-Economic Trading in
Bitcoin.'' Presentation to the U.S. Securities and Exchange Commission,
File SR-NYSEArca-2019-01, March 2019.

Blockchain.com (2026). ``Total Circulating Bitcoin.''
blockchain.com/charts/total-bitcoins. Accessed April 2026.

Bradbury, D. (2014, November 23). ``How Dangerous is Satoshi Nakamoto?''
\emph{CoinDesk}.
coindesk.com/markets/2014/11/23/how-dangerous-is-satoshi-nakamoto.

BtcDrak, M. Friedenbach, and E. Lombrozo (2015). ``BIP 112:
CHECKSEQUENCEVERIFY.'' Bitcoin Improvement Proposals.
github.com/bitcoin/bips/blob/master/bip-0112.mediawiki.

Carreyrou, J., with D. Freedman (2026, April 8). ``My Quest to Solve
Bitcoin's Great Mystery.'' \emph{The New York Times}.
nytimes.com/2026/04/08/business/bitcoin-satoshi-nakamoto-identity-adam-back.html.

Center for Art Law (2018, March 28). ``Blockage Discounts and Artists'
Estates: The De Kooning Post-Mortem.''
itsartlaw.org/case-review/blockage-discounts-and-artists-estates-the-de-kooning-post-mortem.

CNBC (2026, April 8). ``Latest investigation into bitcoin founder ties
identity to Blockstream CEO Adam Back.''
cnbc.com/2026/04/08/latest-investigation-of-bitcoin-founder-ties-identity-to-blockstream-ceo-adam-back.html.

CoinDesk (2024, July 9). ``It's Not Germany Selling Bitcoin. It's One of
Its States and It Has No Choice.'' coindesk.com.

CoinDesk (2025, July 25). ``Bitcoin Rebounds After Galaxy Completes Sale
of \$9B BTC From Satoshi-Era Whale.'' coindesk.com.

Donier, J., and J. Bonart (2015). ``A Million Metaorder Analysis of
Market Impact on the Bitcoin.'' \emph{Market Microstructure and
Liquidity}, 1(2), 1550008. DOI 10.1142/S2382626615500082.

\emph{Estate of David Smith v. Commissioner}, 57 T.C. 650 (1972), aff'd
510 F.2d 479 (2d Cir. 1975).

\emph{Estate of O'Keeffe v. Commissioner}, T.C. Memo 1992-210.

Farmer, J. D., A. Gerig, F. Lillo, and H. Waelbroeck (2013). ``How
efficiency shapes market impact.'' \emph{Quantitative Finance}, 13(11):
1743-1758. DOI 10.1080/14697688.2013.848464.

Glassnode (2023). ``Hodled or Lost Coins'' (BTC on-chain metric).
studio.glassnode.com/metrics?a=BTC\&m=indicators.HodledLostCoins. Metric
reached approximately 7.80 million BTC in August 2023.

Holderness, C. G. (2003). ``A Survey of Blockholders and Corporate
Control.'' \emph{FRBNY Economic Policy Review}, 9(1): 51-64.

Internal Revenue Code § 1014, 26 U.S.C. § 1014 (Basis of property
acquired from a decedent).

IRS Publication 551 (2025). ``Basis of Assets.'' Internal Revenue
Service.

Jalan, A., R. Matkovskyy, and A. Urquhart (2022). ``Demand elasticities
of Bitcoin and Ethereum.'' \emph{Economics Letters}, 220, 110877. DOI
10.1016/j.econlet.2022.110877.

Karantias, K., A. Kiayias, and D. Zindros (2020). ``Proof-of-Burn.'' In
\emph{Financial Cryptography and Data Security 2020 (FC 2020)}, Lecture
Notes in Computer Science 12059, pp.~523-540. Springer. DOI
10.1007/978-3-030-51280-4\_28.

Komlo, C., and I. Goldberg (2021). ``FROST: Flexible Round-Optimized
Schnorr Threshold Signatures.'' In \emph{Selected Areas in Cryptography
-- SAC 2020}, Lecture Notes in Computer Science 12804, pp.~34-65.
Springer. DOI 10.1007/978-3-030-81652-0\_2.

Kyle, A. S. (1985). ``Continuous Auctions and Insider Trading.''
\emph{Econometrica}, 53(6): 1315-1335. DOI 10.2307/1913210.

Lerner, S. D. (2013, April 17). ``The Well Deserved Fortune of Satoshi
Nakamoto, Bitcoin creator, Visionary and Genius.'' \emph{Bitslog}.
bitslog.com/2013/04/17/the-well-deserved-fortune-of-satoshi-nakamoto.

Lerner, S. D. (2020, August 31). ``Protection Over Profit: What Early
Mining Patterns Suggest About Bitcoin's Inventor.'' \emph{CoinDesk}.
coindesk.com/tech/2020/08/31/protection-over-profit-what-early-mining-patterns-suggest-about-bitcoins-inventor.

Makarov, I., and A. Schoar (2021). ``Blockchain Analysis of the Bitcoin
Market.'' NBER Working Paper No.~29396, National Bureau of Economic
Research (working paper; also LSE Financial Markets Group discussion
paper). DOI 10.3386/w29396.

Nakamoto, S. (2008). ``Bitcoin: A Peer-to-Peer Electronic Cash System.''
bitcoin.org/bitcoin.pdf.

Nick, J., T. Ruffing, and Y. Seurin (2021). ``MuSig2: Simple Two-Round
Schnorr Multi-Signatures.'' In \emph{Advances in Cryptology -- CRYPTO
2021}, Lecture Notes in Computer Science 12825, pp.~189-221. Springer.
DOI 10.1007/978-3-030-84242-0\_8.

Roberts, J. J., and N. Rapp (2017, November 25). ``Exclusive: Nearly 4
Million Bitcoins Lost Forever, New Study Says.'' \emph{Fortune}.
fortune.com/2017/11/25/lost-bitcoins.

Rule 10b5-1, 17 C.F.R. § 240.10b5-1 (Trading on the basis of material
nonpublic information in insider trading cases). U.S. Securities and
Exchange Commission.

Said, E. (2022). ``Market Impact: Empirical Evidence, Theory and
Practice.'' arXiv:2205.07385 {[}q-fin.TR{]}. DOI
10.48550/arXiv.2205.07385.

Shamir, A. (1979). ``How to Share a Secret.'' \emph{Communications of
the ACM}, 22(11): 612-613. DOI 10.1145/359168.359176.

Soros, G. (1987). \emph{The Alchemy of Finance}. Simon and Schuster.

Soros, G. (2013). ``Fallibility, reflexivity, and the human uncertainty
principle.'' \emph{Journal of Economic Methodology}, 20(4): 309-329. DOI
10.1080/1350178X.2013.859415.

Spence, A. M. (1973). ``Job Market Signaling.'' \emph{Quarterly Journal
of Economics}, 87(3): 355-374. DOI 10.2307/1882010.

Sward, A., I. Vecna, and F. Stonedahl (2018). ``Data Insertion in
Bitcoin's Blockchain.'' \emph{Ledger}, 3. DOI 10.5195/ledger.2018.101.

Todd, P. (2014). ``BIP 65: OP\_CHECKLOCKTIMEVERIFY.'' Bitcoin
Improvement Proposals.
github.com/bitcoin/bips/blob/master/bip-0065.mediawiki.

\vspace{0.6em}

\end{hangingrefs}

\hypertarget{appendix-a.-scenario-arithmetic-for-the-track-1-absorption-bound}{%
\subsection{Appendix A. Scenario arithmetic for the Track 1 absorption
bound}\label{appendix-a.-scenario-arithmetic-for-the-track-1-absorption-bound}}

This appendix gives three scenarios bounding the cumulative price impact
of a patient OTC sale of the 1.148 million BTC Patoshi position,
relative to the no-sale case. The analysis is stylized and
partial-equilibrium, and the scenarios are illustrative bounds under
stated inputs, not point forecasts and not calibrated estimates. Their
job is to discipline Section 3's qualitative claim that the mechanical
bear case is bounded, to make the sensitivity to each assumption
explicit, and to cross-check the result against an independent method
(A.8).

\hypertarget{a.1-common-parameters}{%
\subsubsection{A.1 Common parameters}\label{a.1-common-parameters}}

Position size: 1.148 million BTC, matching the Patoshi cluster
identified in Lerner (2013, 2020).

Lost coins: the most-cited segmentation study estimates 2.78 to 3.79
million BTC permanently lost as of 2017 (Chainalysis, reported in
Roberts and Rapp, 2017); since losses accumulate, we use the high end as
the central assumption. Dormancy-based upper envelopes used in on-chain
research run much higher but deliberately over-count intentional
holders, and they include the Patoshi coins themselves. For
peer-reviewed active-supply analysis see Ambrosia, Dorrell and Stockwell
(2024). Because a larger lost-coins figure thins the float and raises
estimated impact, A.6 reports the bound under the adverse envelope as
well as the central case.

Freely circulating float F: mined supply minus lost coins minus the
Patoshi position, about \(20.01 - 3.79 - 1.148 \approx 15.1\) million
BTC under the central assumption. So P/F \(\approx\) 7.6 percent. The
position is excluded from the float so that its sale is counted once, as
added supply, rather than as both existing float and added supply.

Reference price: 80,000 USD per BTC, roughly the late-April 2026 level
(bitcoin traded near 76,000 to 78,000 USD over April 24 to 30, 2026).
This is a round figure for scaling, not a dated quote, and all dollar
figures scale linearly with it. It is not the 62,600 USD level of April
8, 2026, the day of the identity reporting discussed in Section 3.2.

Demand elasticity \(\varepsilon_D\) is defined so that a permanent
supply increase of fraction x produces a price change of about
\((1+x)^{-1/\varepsilon_D}-1\). No published study estimates this
structural parameter for bitcoin: the closest peer-reviewed work, Jalan,
Matkovskyy and Urquhart (2022), estimates dynamic demand elasticities
with respect to price, transaction fees, and energy cost, and shows
bitcoin demand elasticity is estimable and non-degenerate, but it does
not identify the price response to a permanent float expansion. The
range \(\varepsilon_D\) = 0.3 (very inelastic) to 1.5 (closer to
equities) is therefore a heuristic sensitivity range, not a confidence
interval, and the paper's quantitative claims rest on triangulating this
arithmetic with the independent methods of A.3 and A.8, not on any
calibration of \(\varepsilon_D\).

Real volume: the participation-rate arithmetic assumes real (non-wash)
spot volume of 10 to 20 billion USD per day. Reported volumes overstate
real ones, historically by a large factor (Bitwise Asset Management,
2019); even under an extreme discount to 1 billion per day, the
program's participation rate is about 2.5 percent, within the range
execution practice treats as low-impact, and the A.8 cross-check prices
the 5-billion case explicitly. The participation rates in A.4 are
computed at the 15-billion-dollar midpoint of this range.

\hypertarget{a.2-partial-equilibrium-impact-by-elasticity}{%
\subsubsection{A.2 Partial-equilibrium impact by
elasticity}\label{a.2-partial-equilibrium-impact-by-elasticity}}

In a partial-equilibrium model with constant demand elasticity, fully
selling the Patoshi position expands freely circulating supply by about
7.6 percent, for a price change relative to the no-sale case of about
\((1.076)^{-1/\varepsilon_D}-1\). The table reports this across the
heuristic elasticity range.

\begin{longtable}[]{@{}ll@{}}
\toprule
Elasticity \(\varepsilon_D\) & Cumulative partial-equilibrium
impact\tabularnewline
\midrule
\endhead
1.5 (equity-like) & approximately \textminus{}4.7 percent\tabularnewline
0.7 (central) & approximately \textminus{}9.9 percent\tabularnewline
0.3 (highly inelastic) & approximately \textminus{}21.6
percent\tabularnewline
\bottomrule
\end{longtable}

Elasticity is the dominant source of variation. Halving
\(\varepsilon_D\) from 0.7 to 0.3 roughly doubles the impact; doubling
it from 0.7 to 1.5 roughly halves it. Demand growth over the sale period
affects the absolute price path but, under the constant-elasticity
log-linear form used here, not the relative impact between the sale and
no-sale cases. The analysis is static: it holds the demand schedule
fixed and treats market-makers as passive. Dynamic adjustment runs both
ways, and we note both. Market-makers who anticipate an announced
program would pre-position ahead of the flow, lowering realized impact.
Conversely, for an asset with no cash-flow floor, demand and offered
float depend on the narrative (Section 3.2): a sale that dented the
upside story would pull demand in and add supply at once, raising
realized impact in bad states. Read the static figures as a central
tendency between these dynamics, capped above by the adverse-stack
calculation of A.6.

\hypertarget{a.3-empirical-anchors-temporary-versus-permanent-impact}{%
\subsubsection{A.3 Empirical anchors: temporary versus permanent
impact}\label{a.3-empirical-anchors-temporary-versus-permanent-impact}}

The partial-equilibrium figure is the permanent part, attributable to
the supply shift. Execution adds a temporary part that the anchors
bound. The Saxony 2024 episode (49,858 BTC in twenty-three days of
publicly tracked tranches; a 13 to 15 percent decline confounded with
the concurrent Mt. Gox distribution announcement and leveraged-long
liquidations; substantial recovery within weeks of the final tranche)
bounds the temporary impact of maximally visible, compressed, public
selling: large while underway, mostly reverting afterward, implying a
permanent part near zero for a 0.3-percent-of-float sale. The U.S.
Marshals auctions (muted to positive responses to institutional
placement) and the July 2025 sale of about 80,000 BTC through a single
OTC desk (a few percent, reverting within days) bound the temporary
impact of disciplined institutional selling. The pattern across anchors,
large-but-reverting under public selling and small under institutional
selling, is the signature of a temporary impact that decays around a
small permanent core, consistent with the near-complete long-run decay
of uninformed-flow impact documented in Donier and Bonart (2015). For
the scenarios we add an execution-friction allowance of 1 to 2 percent
(disciplined OTC), 2 to 3 percent (base), or 3 to 5 percent (mixed
public selling) to the permanent part.

\hypertarget{a.4-three-scenarios}{%
\subsubsection{A.4 Three scenarios}\label{a.4-three-scenarios}}

\emph{Scenario A (conservative).} \(\varepsilon_D\) = 1.5, disciplined
OTC execution, 12-year horizon, participation rate approximately 0.14
percent of real daily spot volume, pace approximately 95,600 BTC per
year. Permanent impact approximately \textminus{}4.7 percent. Execution
friction approximately 1 to 2 percent. Total cumulative impact relative
to counterfactual: approximately \textminus{}6 to \textminus{}7 percent.

\emph{Scenario B (base).} \(\varepsilon_D\) = 0.7, disciplined OTC
execution, 10-year horizon, participation rate approximately 0.17
percent of real daily spot volume, pace approximately 114,800 BTC per
year. Permanent impact approximately \textminus{}9.9 percent. Execution
friction approximately 2 to 3 percent. Total cumulative impact:
approximately \textminus{}12 to \textminus{}13 percent.

\emph{Scenario C (aggressive).} \(\varepsilon_D\) = 0.3, mixed execution
quality (partial public venue), 5-year horizon, participation rate
approximately 0.34 percent of real daily spot volume, pace approximately
229,600 BTC per year. Permanent impact approximately \textminus{}21.6
percent. Execution friction approximately 3 to 5 percent. Total
cumulative impact: approximately \textminus{}25 to \textminus{}27
percent.

\hypertarget{a.5-sensitivity}{%
\subsubsection{A.5 Sensitivity}\label{a.5-sensitivity}}

The dominant sensitivity is to elasticity; the second is to the
lost-coins assumption (A.6). Scenario C approaches the Saxony episode's
headline decline under its adverse stack, with the key difference that
the Saxony decline largely reversed while Scenario C's permanent part
would not; Scenario A approaches the institutional-execution anchors;
Scenario B sits between them.

\hypertarget{a.6-robustness-and-takeaway}{%
\subsubsection{A.6 Robustness and
takeaway}\label{a.6-robustness-and-takeaway}}

Under the adverse lost-coins envelope (treating the dormancy-based upper
bound of roughly 7.8 million BTC, minus the Patoshi coins it includes,
as genuinely lost; this is the peak of Glassnode's ``Hodled or Lost
Coins'' metric, about 7.80 million BTC in August 2023), F falls to about
12.2 million and P/F rises to about 9.4 percent; the permanent impact
becomes about \textminus{}5.8 percent (\(\varepsilon_D\) = 1.5),
\textminus{}12.0 percent (0.7), and \textminus{}25.9 percent (0.3), and
the aggressive total reaches about \textminus{}29 to \textminus{}31
percent. So the existential-tail framing fails even under the
assumptions most favorable to it: envelope losses, the most inelastic
proposed demand, and adverse execution together bound the cumulative
impact near \textminus{}30 percent relative to the no-sale case, a
material drawdown but not a collapse. Under the central lost-coins
assumption the base case stays about \textminus{}10 to \textminus{}13
percent; under the adverse envelope the same central-elasticity case
rises to roughly \textminus{}14 to \textminus{}15 percent after
execution friction. No plausible calibration within this mechanical
supply model supports valuing the disposition at a large fraction of the
pre-event price. The mechanical bear case is bounded by the arithmetic.

\hypertarget{a.7-the-consistency-ledger-formally}{%
\subsubsection{A.7 The consistency ledger,
formally}\label{a.7-the-consistency-ledger-formally}}

Let F and P be as in A.1, let \(\pi\in[0,1]\) be the expected fraction
of P the market currently prices as eventually returning to float, and
let \(u\geq 0\) be a residual discount for the unresolved state. With
constant-elasticity demand \(p(S)\propto S^{-1/\varepsilon}\), the
burn-confirmed price is p(F), the sale-confirmed price is p(F + P), and
the current price is approximately \(p(F+\pi P)(1-u)\). Define
\(T=(1/\varepsilon)\ln(1+P/F)\), the total log spread between the two
confirmed states: \(T\approx 0.049\) (\(\varepsilon=1.5\)), \(0.104\)
(\(\varepsilon=0.7\)), \(0.243\) (\(\varepsilon=0.3\)). The
burn-confirmed log upside from the current price is
\((1/\varepsilon)\ln(1+\pi P/F)-\ln(1-u)\), and the sale-confirmed log
downside is \((1/\varepsilon)\ln((F+P)/(F+\pi P))+\ln(1-u)\). To first
order these are \(\pi T+u\) and \((1-\pi)T-u\). The two supply
components, (1/\(\varepsilon\)) ln(1 + \(\pi\)P/F) and
(1/\(\varepsilon\)) ln((F + P)/(F + \(\pi\)P)), sum exactly to T; this
is the complementarity invoked in Section 3.4, namely that the
confirmation upside and the unanticipated-sale downside are shares of
one bounded supply spread. The unresolved-state discount u is handled
separately: if it resolves on either outcome, it adds to the burn upside
and offsets the sale downside by the same amount, so it cancels in the
sum and the two total returns also sum to T. Note that with u
\textgreater{} 0 the current price sits below the interior point
\(p(F+\pi P)\), and a confirmed sale can even raise the price if u
exceeds the supply loss \((1-\pi)T\). An aliveness reveal without coin
movement reprices only \(\Delta\pi T\). A.2 through A.6 report the
\(\pi\) = 0, u = 0 polar case and are therefore upper bounds on the
unanticipated supply component of a sale.

\hypertarget{a.8-cross-check-square-root-law}{%
\subsubsection{A.8 Cross-check: square-root
law}\label{a.8-cross-check-square-root-law}}

As an independent method, treat each year of a ten-year program as a
metaorder of about 114,800 BTC and apply the square-root impact law
\(I\approx Y\cdot\sigma\cdot\sqrt{Q/V}\), which Donier and Bonart (2015)
document on bitcoin across four decades of order size. With annualized
volatility of 45 to 60 percent and real spot volume of 5 to 20 billion
USD per day, and using annual volume (V = 365 times daily volume) to
match the annual metaorder Q, peak impact per annual metaorder is about
1.6 to 4.3 percent, of which the post-completion component, taken as
roughly two-thirds of peak under the fair-pricing condition (Farmer,
Gerig, Lillo and Waelbroeck, 2013), implies 1.1 to 2.8 percent per
annual metaorder before allowing for later decay, overlap, or saturation
across successive same-direction metaorders. A naive ten-year sum of 11
to 28 percent overstates the cumulative impact, since it ignores the
near-complete long-run decay of uninformed-flow impact documented in the
same study and the saturation of successive same-direction metaorders.
Read as an order-of-magnitude bound, a method sharing no parameters with
the elasticity arithmetic lands in the same range as A.4: single digits
to the mid-twenties, with a central tendency near ten to fifteen
percent. The convergence of the two methods, with the episode anchors of
A.3, is why Section 3 calls the bound robust despite the lack of a
calibrated elasticity.

\vspace{0.6em}

\hypertarget{appendix-b.-the-mechanics-of-provable-destruction}{%
\subsection{Appendix B. The mechanics of provable
destruction}\label{appendix-b.-the-mechanics-of-provable-destruction}}

An OP\_RETURN output is a transaction output whose locking script starts
with the OP\_RETURN opcode, which makes script evaluation fail
unconditionally. Because the script fails whenever it is evaluated, no
unlocking data of any kind can make the output spendable, so any value
assigned to it is provably unspendable, and nodes drop these outputs
from the UTXO set rather than carry them as spendable. Relay of
transactions with one small OP\_RETURN output was standardized in
Bitcoin Core 0.9.0 in March 2014 (Bitcoin Wiki, 2024; Sward, Vecna and
Stonedahl, 2018, give the comprehensive survey). A burn is a transaction
that sends value to such an output. It needs no data payload, no special
infrastructure, and no counterparty, and anyone can verify it
immediately and without trust. Burning the whole Patoshi cluster would
be a matter of ordinary standard transactions, differing from everyday
OP\_RETURN use, which usually attaches negligible value to a
data-carrying output, only in the value sent. The formal cryptographic
treatment of burn protocols, including their security properties, is in
Karantias, Kiayias and Zindros (2020).

Three grades of destruction should be distinguished, because they differ
in verifiability even when their supply effect is identical. First, the
provable burn just described: a cryptographically verifiable
subtraction, effective and confirmable in one block. Second, transfers
to ``burn addresses'' with no known private key, such as the address
used in the Counterparty proof-of-burn of January 2014, the largest
deliberate burn in bitcoin's history at slightly over 2,100 BTC:
unspendable with overwhelming probability, but resting on the absence of
a known key rather than on script semantics. Third, key destruction: a
physically equivalent subtraction that no outsider can ever verify,
since you cannot prove that no key copies exist. The ledger of Section
3.4 prices the difference. A provable burn moves the market's residual
sale probability to zero credibly and at once, delivering the full
upside in one block. Key destruction, however sincere, delivers it only
gradually, as the market updates on continued dormancy, and in principle
never fully. A Patoshi burn would be roughly five hundred times the
largest burn ever done: unprecedented in scale and mechanically trivial.

One current development is worth noting. OP\_RETURN relay policy became
contested in 2025, when Bitcoin Core version 30 (October 2025) raised
the default data-carrier limit from 83 bytes to 100,000 bytes and
allowed multiple OP\_RETURN outputs per transaction (Bitcoin Core,
2025), prompting a visible split in node-policy preferences between
Bitcoin Core and the more conservative Bitcoin Knots distribution. The
dispute is about relaying \emph{data}, not about the treatment of
\emph{value}: the consensus rule that value sent to an OP\_RETURN output
is unspendable has been stable since introduction and is untouched by
either side of the debate. A value burn needs no data payload, so it is
insulated from the controversy. The destruction option that Section 7
calls unique to bitcoin rests on this consensus-level property, not on
relay policy.

\end{document}